\documentclass[onecolumn, revtex-4, tighten]{openjournal}

\usepackage{graphicx}
\usepackage[utf8]{inputenc} % allow utf-8 input
\usepackage[T1]{fontenc}    % use 8-bit T1 fonts
\usepackage[hidelinks]{hyperref}       % hyperlinks
\usepackage{url}            % simple URL typesetting
\usepackage{booktabs}       % professional-quality tables
\usepackage{amsfonts}       % blackboard math symbols
\usepackage{nicefrac}       % compact symbols for 1/2, etc.
\usepackage{microtype}      % microtypography
\usepackage{amsmath, amssymb}
\allowdisplaybreaks

\newcommand{\DM}{{\rm DM}}
\newcommand{\dd}{{\rm d}}

\begin{document}

\title{Pad\'e Approximants for cosmic Dispersion Measures}

\author{Marios Kalomenopoulos$^{1, 2}$ \href{https://orcid.org/0000-0001-6677-949X}{\includegraphics[scale=0.6]{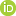}}}
\author{Jiaming Zhuge$^{1, 2}$ \href{https://orcid.org/0000-0001-8114-3094}{\includegraphics[scale=0.6]{Figures/ORCIDiD_icon.png}}}

\affiliation{$^{1}$ Nevada Center for Astrophysics, University of Nevada, Las Vegas, NV 89154, USA\\ 
  $^{2}$ Department of Physics and Astronomy, University of Nevada, Las Vegas, NV 89154, USA}

\thanks{$^*$E-mail: marios.kalomenopoulos@gmail.com, jiaming.zhuge@unlv.edu}

\begin{abstract}

Fast Radio Bursts (FRBs) have become an indispensable tool for studying the ``missing baryons'', the Universe's ionisation properties, as well as the cosmological parameters. This is achieved by analysing the diffuse dispersion measure ($\DM_{\rm diff}$) of FRBs as a function of redshift. However, the rapidly increasing data size requests more and more computational resources. In this work, we first develop an accelerated method for any cosmic dispersion measure by deriving an analytical approximation formula for flat, $\Lambda$CDM and $w$CDM universes. Focusing on FRBs, we show that our approximation works well for the ranges $0.01 \leq z \leq 2$, $0.2 \leq \Omega_m \leq 1.0$ and $-3.0 \leq w \leq -0.5$, with relative error to a numerically evaluated $\DM_{\rm diff}$ always smaller than $3.5 \%$ (in the worst case scenario). This error remains below observationally relevant $\DM$ scatter and is especially small near the concordance $\Lambda$CDM cosmology. Additionally, we perform a cosmological analysis of simulated FRB data and show that our approximation gives robust and unbiased results, even when applied in regions of parameter space where its relative error becomes larger than $3\%$. Finally, the approximation is more than $15$ ($2$) times faster than the numerical solution of $\Lambda$CDM ($w$CDM), and can reach a timing improvement of a factor of $25$ when used in an MCMC cosmological inference.

\end{abstract}


\maketitle

\section{Introduction}\label{sec:intro}

The dispersion measure (DM) of a radio source reflects the presence of ionised gas in the observer's line of sight. If the source is at a cosmological distance, the cosmic DM can reveal the ``missing baryons'' in the Universe, as well as cosmological and/or ionisation parameters. Among these cosmic radio sources, fast radio bursts (FRBs) \citep[e.g.][]{Lorimer_burst_2007, Cordes_Chatterjee_FRB_review_2019, Petroff_FRB_review_2019, Petroff_FRB_review_2022, Wu_Wang_FRB_cosmo_review_2024, Glowacki_Lee_FRB_cosmo_Review_2024} have become a unique probe due to their high energy pulses and cosmological origins\footnote{Only one FRB, FRB 20200428, has been connected to a galactic magnetar \citep{2020Nature_galactic_FRB}.}. At the same time, they offer multiple opportunities for multimessenger astrophysics, if they can be associated with other high-energy astrophysical phenomena \citep{MMA_FRBS_Bing_2024, FRBs_GWs_cosmo_assoc}.

With the improving observation and localization ability of FRBs, their statistics and cosmological studies are developing rapidly \citep{Macquart+_2020, Connor_et_al_2025_missing_baryons, FRBcosmo_localised_Zhuge_et_al_2026}. However, with the increasing data size, FRB cosmology is also becoming a likelihood and hierarchical inference problem since increasing computational resources would be needed, which additionally can hinder new researchers with limited computational infrastructure to join this area. 

For this reason, a fast and accurate calculation of the theoretical, cosmic dispersion measure of FRBs would be crucial. Pad\'e approximants \citep{Pade_paper} offer a precise and efficient alternative to the standard numerical integrations that are needed, with many cosmological applications already. \citet{Adachi_Kasai_dL_Pade_2012} first developed a Pad\'e approximation to analytically calculate the luminosity distance in a $\Lambda$CDM cosmology, showing that it outperforms previous similar approximations \citep{Pen_dL_1999, Wickramasinghe_Ukwatta_dL_2010}. Then, \citet{Wei_Yan_Zhou_Pade_wCDM_2014} derived a more general Pad\'e approximation for the luminosity distance under a $w$CDM cosmology.

In this work, we derive for the first time, as far as we are aware, a Pad\'e approximation for cosmic DMs. In Section \ref{sec:Dispersion_Measure} we describe the DM cosmic integral. We then develop our approximations considering both $\Lambda$CDM and $w$CDM cosmologies, in Sections \ref{sec:Pade_for_DM_LCDM} and \ref{sec:Pade_for_DM_wCDM} respectively. In Section \ref{sec:accuracy_timing_performance}, we quantify the accuracy and time performance of the approximation, while in Section \ref{sec:cosmo_performance} we assess its performance in inferring cosmological parameters. Finally, we summarise our results in Section \ref{sec:conclusion}.

\section{Cosmic Dispersion Measure for FRBs}\label{sec:Dispersion_Measure}

FRBs are bursts of radio emission that last for the order of milliseconds. As they travel through the Universe the different frequencies of the signal travel with different speeds and arrive at different times. For two photons travelling from a cosmological distance of redshift $z$, with frequencies $\nu_1$ and $\nu_2$ much greater than the plasma frequency $\omega_p$, the time difference of arrival measured by an observer is \citep{1979Radiative_processes_in_astrophysics, FRB_cosmo_DM_Deng_Zhang_2014}:
\begin{equation}
    \Delta t=\frac{e^2}{2\pi m_e c}\left(\frac{1}{\nu_1^2}-\frac{1}{\nu_2^2}\right)\int_0^L \frac{n_e(z)}{1+z}\dd{l},
\end{equation}
where $e, m_e, n_e$ are the charge, mass and number density of electrons, $c$ is the speed of light and $L$ is the proper distance from the FRB source to the observer. These time delays are quantified by the dispersion measure of the pulse, defined as
\begin{equation}
    \DM=\int_0^L \frac{n_e(z)}{1+z}\dd{l}.
\end{equation}

During its propagation through the Universe, the FRB pulse passes through various environments, whose dispersion measures is usually analysed separately \citep[e.g.][]{FRB_DM_break_Thornton_et_al_2013, FRB_cosmo_DM_Deng_Zhang_2014, FRB_DM_break_Prochaska_Zheng_2019}. In this work, we are focusing on one such contribution, $\DM_{\rm diff}$.

The $\DM_{\rm diff}$ measures the contributions of diffused electrons along the line-of-sight, drawn from the intergalactic medium (IGM) and intervening cosmic halos. This has the clearest connection to cosmology, since a mean diffuse $\DM$ can be derived theoretically and depends on the various cosmological parameters. Assuming a homogeneous distribution and ionization of baryons, with ionized fraction $\chi(z)$ \citep{DMdiff_Ioka_2003, DMdiff_Inoue_2004, FRB_cosmo_DM_Deng_Zhang_2014} one has

\begin{equation}
    \langle{\DM_{\rm diff}}\rangle 
    (z) = \frac{3cH_0\Omega_b }{8\pi G m_p}\int_0^z \frac{ f_{\rm diff}(z')\chi(z')(1+z')}{E(z', \Omega_i)}\dd{z'},
\end{equation}

with
\begin{equation}
    \chi(z) = Y_{\rm H} X_{e, \textrm{H}}(z) + \frac{1}{2}Y_{\rm He} X_{e, \textrm{He}}(z).
\end{equation}

The different parameters correspond to: $H_0$ the Hubble parameter, $\Omega_b$ the density of baryons in the Universe, $G$ Newton's gravitational constant, $m_p$ the mass of the proton, $\Omega_m$ the density of matter in the Universe, $\Omega_\Lambda$ the density due to a cosmological constant, and where $E(z, \Omega_i)$ is the dimensionless Hubble parameter which will be defined for each specific case below (Sections \ref{sec:Pade_for_DM_LCDM} and \ref{sec:Pade_for_DM_wCDM}). 

Additionally, $f_{\rm diff}$ introduces astrophysical effects, quantifying the fraction of diffuse baryons in both the intergalactic medium and halos, and can in principle depend on redshift. Following, observational studies that have constrained $f_{\rm diff} \sim 0.84$ with no evidence of redshift dependence \citep{Li_et_al_fdiff_2020}, we take it here as a constant and take it out of the integral. $\chi(z)$ represents the fraction of ionized electrons over the baryons in the Universe. Its components correspond to the mass fraction in the Universe of hydrogen $\textrm{H},\ Y_{\rm H}$ or helium $\textrm{He}, Y_{\rm He}$ and to the ionization fraction of each element. For low redshifts $(z<3)$, the Universe is ionized, and $\chi(z)$ tends to a constant value $\chi(z) = \chi_{7/8} \sim 7/8$ \citep{Fukugita_Hogan_Peebles_1998, FRB_cosmo_DM_Deng_Zhang_2014}. As a result $\DM_{\rm diff}$ becomes

\begin{align}\label{eq:DM_diff}
    \langle{\DM_{\rm diff}}\rangle 
    (z) &= \frac{3cH_0\Omega_b f_{\rm diff}}{8\pi G m_p} \chi_{7/8} \int_0^z \frac{1+z'}{E(z', \Omega_i)}\dd{z'} \nonumber \\
    &= \DM_{\rm diff}^c \int_0^z \frac{1+z'}{E(z', \Omega_i)}\dd{z'},
\end{align}

where in $\DM_{\rm diff}^c$ we have collected all the cosmological and astrophysical parameters that do not affect the $z$ integral.

\section{Method}
\label{sec:method}

Our main goal is to find a good approximation to the integral in eq. (\ref{eq:DM_diff}). Our general approach would be to expand it, in a suitable limit, to a power series $\Sigma a_n x^n$ and then fit the series to a Pad\'e approximant of the form

\begin{equation}\label{eq:Pade_series}
    P_M^N(z) = \frac{\sum A_n x^n}{\sum B_n x^n}.
\end{equation}

The power of the Pad\'e approximant is that it can resemble the original function which is expanded to a power series, even if the latter is divergent \citep{Bender_Orszag_1999}.

\subsection{Pad\'e Approximant for $\DM_{\rm diff}$ - Flat $\Lambda$CDM}\label{sec:Pade_for_DM_LCDM}

We obtain our approximation, following closely the equivalent derivation of \cite{Adachi_Kasai_dL_Pade_2012} for the luminosity distance in flat $\Lambda$CDM cosmologies. We start from eq. (\ref{eq:DM_diff}):

\begin{align}\label{eq:Phi_alpha}
    \DM_{\rm diff} =& \DM_{\rm diff}^c \int_0^z \frac{ (1+z) dz}{\sqrt{\Omega_m(1+z)^3+(1-\Omega_m)}} \nonumber \\ 
    =& \DM_{\rm diff}^c \int_a^1 \frac{da}{\sqrt{\Omega_m a^3+(1-\Omega_m) a^6}},\quad \textrm{with} \quad a=1/(1+z).
\end{align}

We define a new function $F(a)$:

\begin{equation}\label{eq:F_alpha}
    F(a) = \int_0^a \sqrt{\frac{\Omega_m}{\Omega_m a^3+(1-\Omega_m) a^6}} da.
\end{equation}
Note that we have introduced an extra $\sqrt{\Omega_m}$ term, which helps simplify the expansion below. This would re-write eq. (\ref{eq:Phi_alpha}) as
\begin{equation}
    \DM_{\rm diff} = \frac{\DM_{\rm diff}^c}{\sqrt{\Omega_m}} \Big( F(1) - F(a) \Big).
\end{equation}

For the next steps, we are going to expand $F(a)$ with respect to $a$, where $a \rightarrow 0$ (large $z$ regime). This expansion results in

\begin{equation}
    F(a) = \frac{1}{\sqrt{a}} \Bigg( -2 - 0.2 x + 0.068x^2 - 0.037x^4 + \cdots  \Bigg),
\end{equation}

with $x = x(a, \Omega_m) = x(z, \Omega_m)$ is defined as
\begin{equation}\label{eq:x_z_Om}
    x(z, \Omega_m) = \frac{1-\Omega_m}{\Omega_m} \cdot \frac{1}{(1+z)^3}.
\end{equation}

We expand $F$ until $\mathcal{O}(x^6)$ inclusive, and calculate the Pad\`e approximant \citep{Bender_Orszag_1999} to order $(N, M) = (3, 3)$:

\begin{equation}
    \frac{1}{\sqrt{a(x)}} \Phi(x) = \frac{1}{\sqrt{a(x)}} \frac{b_0+b_1x+b_2x^2+b_3x^3}{1.0+c_1x+c_2x^2+c_3x^3}.
\end{equation}

The coefficients are

\begin{align}
    &b_0 = -2.0, \\
    &b_1 = -2.85592665, \\
    &b_2 = -1.0945641, \\
    &b_3 = -0.0913347, \\
    &c_1 = 1.32796333, \\    
    &c_2 = 0.44857662, \\    
    &c_3 = 0.02769881.
\end{align}

Bringing everything together and returning to $(z, \Omega_m)$ notation, we have the analytical formula:

\begin{equation}
    \DM_{\rm diff} = \frac{\DM_{\rm diff}^c}{\sqrt{\Omega_m}}  \Bigg( \Phi \big[ x(0, \Omega_m) \big] - \sqrt{1+z} \cdot \Phi \big[ x(z, \Omega_m) \big]  \Bigg).
\end{equation}

We end this section by noting that in the case of flat $\Lambda$CDM cosmologies, an analytical expression based on special functions exists (see App. \ref{App:Hypergeometric_function}). However, in section \ref{sec:accuracy_timing_performance}, we show that our approximation is competitively accurate for all current and near future FRB observations. Additionally, it is vitally faster when considering complex FRB likelihood calculations.

\subsection{Pad\'e Approximant for $\DM_{\rm diff}$ - Flat $w$CDM}\label{sec:Pade_for_DM_wCDM}

We obtain our approximation, following closely the equivalent derivation of \cite{Wei_Yan_Zhou_Pade_wCDM_2014} for the luminosity distance in flat $w$CDM cosmologies. Here $w$ is the general factor in the equation of state parameterization for dark energy (DE) connecting its pressure $p$ and density $\rho$ as $p=w \rho c^2$ \citep{Hobson_et_al_GRbook}. This modifies the denominator in the different distance elements \citep{Linder_DE_eos_2003, Frieman_Turner_Huterer_DE_2008}, hence the diffuse $\DM$ is now written as:

\begin{align}\label{eq:Phi_alpha_wCDM}
    \DM_{\rm diff}^{w{\rm CDM}} =& \DM_{\rm diff}^c \int_0^z \frac{ (1+z) dz}{\sqrt{\Omega_m(1+z)^3+(1-\Omega_m)(1+z)^{3(1+w)}}} \nonumber \\ 
    =& \DM_{\rm diff}^c \int_a^1 \frac{da}{\sqrt{\Omega_m a^3+(1-\Omega_m) a^{3(1-w)}}},\quad \textrm{with} \quad a=1/(1+z).
\end{align}

For the following analysis, we consider $w$ negative, and specifically $w<-1/6$, to avoid divergences in the denominator. This range is consistent with most standard DE models \citep{DE_EoS_2024}. We define a new function $\tilde{F}(a)$:
\begin{align}\label{eq:F_alpha_wCDM}
    \tilde{F}(a) &= \frac{1}{\sqrt{\Omega_m}}\int_0^a \frac{1}{\sqrt{a^3+\frac{1-\Omega_m}{\Omega_m} a^3 a^{-3w}}} da \Rightarrow \nonumber \\
    \tilde{F}(\tilde{x}) &= - \frac{1}{\sqrt{\Omega_m}} \cdot \frac{1}{3w} \int_0^{\tilde{x}} \left( \frac{s}{\tilde{x}}\right)^{-1/6w} \cdot \frac{1}{\tilde{x}} \cdot \frac{d \tilde{x}}{\sqrt{1+\tilde{x}}},
\end{align}

where we have introduced $s=(1-\Omega_m)/\Omega_m$ and $\tilde{x} = s a^{-3w}$. Note that setting $w=-1$ returns to eq. (\ref{eq:F_alpha}). Using $\tilde{F}$ would re-write eq. (\ref{eq:Phi_alpha_wCDM}) as:
\begin{equation}
    \DM_{\rm diff}^{w{\rm CDM}} = \DM_{\rm diff}^c \Big( \tilde{F}(1) - \tilde{F}(a) \Big).
\end{equation}

Similarly with section \ref{sec:Pade_for_DM_LCDM}, we expand $\tilde{F}(a)$ with respect to $a$, where $a \rightarrow 0$ (large $z$ regime) and then calculate the Pad\`e approximant to order $(N, M) = (3, 3)$:

\begin{equation}\label{eq:Pade_expansion_wCDM}
    \Phi^{w{\rm CDM}}(\tilde{x}) = \frac{b_0+b_1\tilde{x}+b_2\tilde{x}^2+b_3\tilde{x}^3}{1.0+c_1\tilde{x}+c_2\tilde{x}^2+c_3\tilde{x}^3},
\end{equation}

with coefficients reported in detail in App. \ref{App:Pade_wCDM_coeff}. Returning to the $(z, \Omega_m)$ notation, we get the analytical formula:

\begin{equation}
    \DM_{\rm diff}^{w{\rm CDM}} = -\frac{\DM_{\rm diff}^c}{3w\sqrt{\Omega_m}}  \Bigg( \Phi^{w{\rm CDM}} \big[ x(0, \Omega_m, w) \big] - \sqrt{1+z} \cdot \Phi^{w{\rm CDM}} \big[ x(z, \Omega_m, w) \big]  \Bigg).
\end{equation}

\section{Accuracy and Timing Performance}\label{sec:accuracy_timing_performance}

To calculate the accuracy of our approximants, we compare the relative error between the numerical and the approximate calculation of the integrals. We define the fractional percentage error as
\begin{equation}
    \Delta E \ [\%] = \frac{|\DM^{\rm num} - \DM^{\rm Pade}|}{\DM^{\rm num}} \cdot 100
\end{equation}

Wherever we use values of the cosmological parameters from $\Lambda$CDM, we base ourselves on \verb|Planck18| \citep{Planck_Cosmo_param_2018}. For example, when testing the $w$CDM approximant for different $w$ values, the matter density parameter is fixed $\Omega_m = \Omega_m^{\Lambda\textrm{CDM}}$, unless stated otherwise.

\begin{figure}[h]
    \centering
    \includegraphics[width=\linewidth]{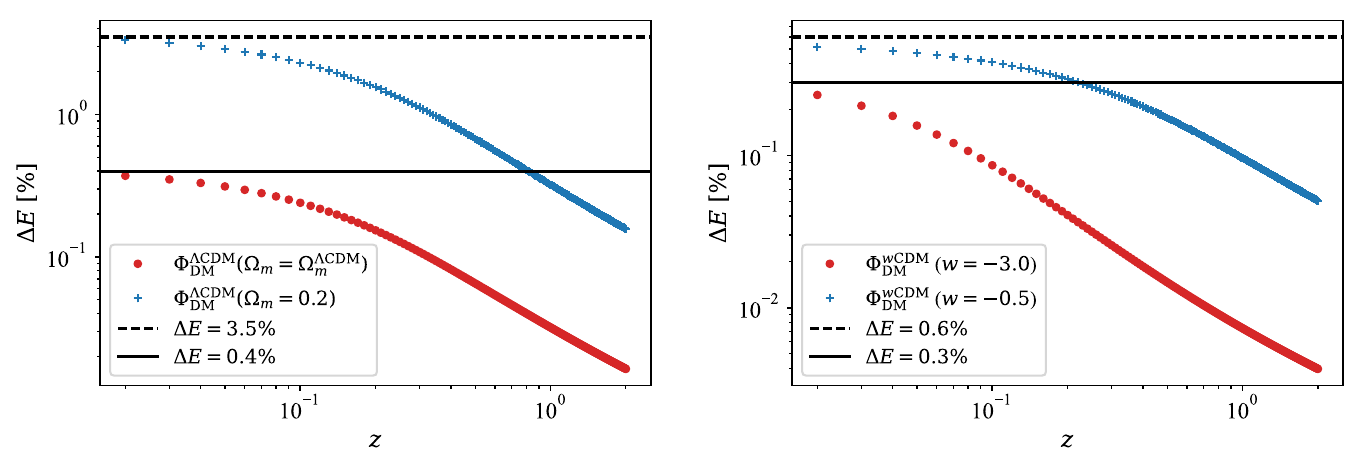}
    \caption{(\textbf{Left}) Fractional error $\Delta E$ for the $\DM$ Pad\'e Approximant in a flat, $\Lambda$CDM universe, for the redshift range $0.01 \leq z \leq 2$, and for two values of $\Omega_m$: ($\Omega_m = 0.2$ - blue, crosses) and  ($\Omega_m = \Omega_m^{\Lambda \textrm{CDM}}$ - red, dots). Increasing $z$ and $\Omega_m$ reduces the fractional error. (\textbf{Right}) Fractional error $\Delta E$ for the $\DM$ Pad\'e Approximant in a flat, $w$CDM universe, for the redshift range $0.01 \leq z \leq 2$, and for two values of $w$: ($w = -0.5$ - blue, crosses) and  ($w = -3.0$ - red, dots). Increasing $z$ and decreasing $w$ reduces the fractional error. Here $\Omega_m = \Omega_m^{\Lambda \textrm{CDM}}$.}
    \label{fig:DeltaE_PhiDM_combined}
\end{figure}

In Figure \ref{fig:DeltaE_PhiDM_combined}, we show the accuracy performance of the Pad\'e approximants for the $\Lambda$CDM and $w$CDM cases. The redshift range is the same for both, $0.01 \leq z \leq 2$. We observe, as expected, that the approximate solutions improve for higher redshifts, and that $\Omega_m$ has a bigger impact on the accuracy, compared to $w$. In general, smaller $\Omega_m$ and more positive $w$ lead to worse results, with $\Omega_m = 0.2$ in the flat, $\Lambda$CDM case yielding the worst\footnote{However, this is still smaller than observational uncertainties of current FRBs' $\DM_{\rm diff}$ (for example, see \cite{Macquart+_2020, Connor_et_al_2025_missing_baryons}).} performance $\Delta E \sim 3.5\ \%$. Table \ref{tab:DE_table_LCDM} gives example values for $\Delta E$ for pairs of $(z, \Omega_m)$. Overall, in the cosmological region most relevant to current data, the approximation error is far below the intrinsic FRB $\DM$ scatter and, as we demonstrate in section \ref{sec:cosmo_performance}, it does not propagate into a visible posterior shift.

\begin{table}[h]
    \centering
    \begin{tabular}{|c|c|c|c|c|c|}
    \hline
        $\Omega_m$ & $z=0.01$ & $z=0.05$ & $z=0.1$ & $z=0.5$ & $z=1.0$ \\
        \hline
        \hline
        $0.2$ & $3.51$ & $2.89$ & $2.31$ & $0.68$ & $0.32$ \\
        $0.3$ & $0.47$ & $0.38$ & $0.29$ & $0.08$ & $0.04$ \\
        $0.4$ & $0.07$ & $0.05$ & $0.04$ & $0.01$ & $0.01$ \\
    \hline
    \end{tabular}
    \caption{$\Delta E \ [\%]$ values for different redshift and $\Omega_m$ values for the flat, $\Lambda$CDM case.}
    \label{tab:DE_table_LCDM}
\end{table}

\begin{figure}[h]
    \centering
    \includegraphics[width=0.85\linewidth]{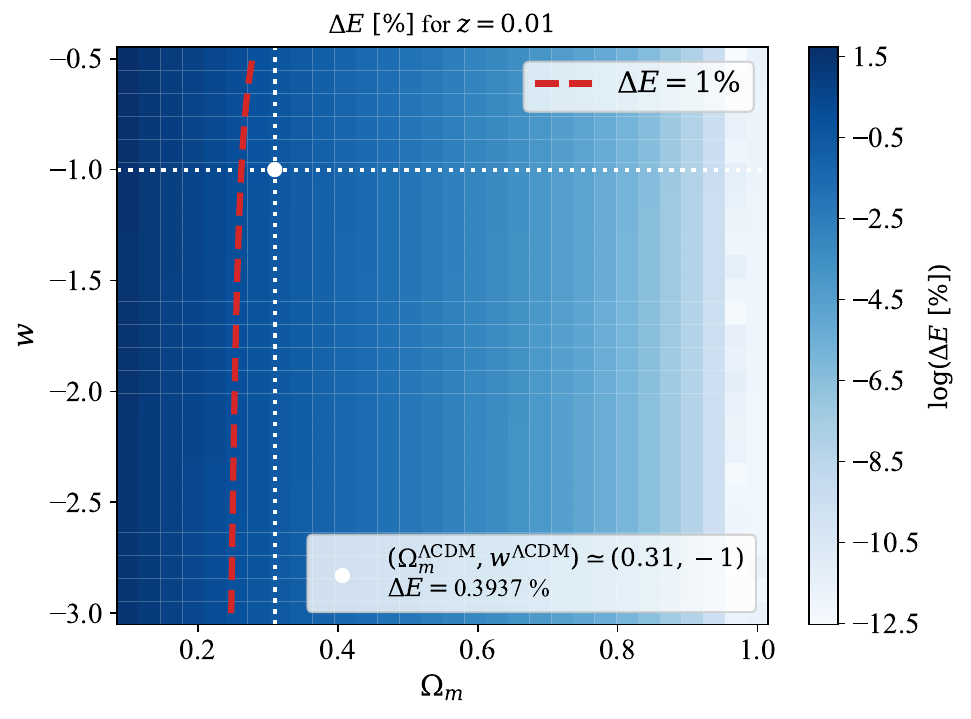}
    \caption{Fractional error $\Delta E$ for the $\DM$ Pad\'e Approximant in a flat, $w$CDM universe, for a fixed redshift $z=0.01$, i.e. the most conservative choice. The other cosmological parameters vary in the ranges: $-3 \leq w \leq -0.5$ and $0.1 \leq \Omega_m \leq 1$. We also denote with a red, dashed line the limit where the fractional error is $\Delta E = 1 \%$, while the white, dotted line corresponds to the cases where the parameters take their $\Lambda$CDM values. We observe that the value of $\Omega_m$ plays the more crucial role in our Pad\'e approximants, compared to $w$.}
    \label{fig:DeltaE_PhiDM_wCDM_2D_plot}
\end{figure}

In Figure \ref{fig:DeltaE_PhiDM_wCDM_2D_plot}, we fix the redshift to the most pessimistic value $z=0.01$, i.e. to the redshift limit where the approximation is anticipated to achieve the worst accuracy. We then compare the accuracy $\Delta E$ for different combinations of $(w, \Omega_m)$. Some example choices are shown in Table \ref{tab:DE_table_wCDM}. The 2D result supplements the 1D analysis of Figure \ref{fig:DeltaE_PhiDM_combined}, showing that for the whole parameter space we studied, the influence of $\Omega_m$ is much more important than $w$.

\begin{table}[h]
    \centering
    \begin{tabular}{|c|c|c|c|}
    \hline
        $w$ & $\Omega_m=0.2$ & $\Omega_m=0.3$ & $\Omega_m=0.4$ \\
        \hline
        \hline
        $-3.0$ & $2.62$ & $0.36$ & $0.05$  \\
        $-1.5$ & $3.1$ & $0.42$ & $0.06$  \\
        $-1.0$ & $3.5$ & $0.47$ & $0.07$  \\
        $-0.5$ & $4.93$ & $0.64$ & $0.09$  \\
    \hline
    \end{tabular}
    \caption{$\Delta E \ [\%]$ values for different $w$ and $\Omega_m$ values for the flat, $w$CDM case, at redshift $z=0.01$.}
    \label{tab:DE_table_wCDM}
\end{table}

\begin{table}[h]
    \centering
    \begin{tabular}{|c|c|c|}
    \hline
        Model & $\Lambda$CDM (Num) & $w$CDM (Num) \\
        \hline
        \hline
        $\Lambda$CDM (Pad\'e) & $\sim 17$ & $-$ \\
        $w$CDM (Pad\'e) & $-$ &  $\sim 2.5$ \\
    \hline
    \end{tabular}
    \caption{Timing improvement $\Delta t = t_{\rm Num}/t_{\rm App}$ values between the numerical solutions and the Pad\'e approximants.}
    \label{tab:Timing_table_Num_Pade}
\end{table}

Finally, in Table \ref{tab:Timing_table_Num_Pade} we compare the timing performance of the approximants versus the numerical integration in \verb|Python|'s \verb|quad| function (The numbers correspond to how much faster is the approximant, which is always true). The biggest improvement is seen in the $\Lambda$CDM case, with around $17$ times faster evaluations. The $w$CDM approximate result still evaluates faster, but with smaller timing improvement. This is because the Pad\'e approximant in this case has more complicated coefficients than the $\Lambda$CDM case (see App. \ref{App:Pade_wCDM_coeff}). However, if one considers the need of repeated calculations of $\DM$ in a bayesian inference scenario the timing improvement will eventually be remarkable. We investigate this further in section \ref{sec:cosmo_performance}.

\section{Cosmological Performance}\label{sec:cosmo_performance}

To assess the impact of our approximation on cosmological studies, we generate $N=50$ simulated FRB events, between redshifts $z \ \epsilon \ [0.25, 2]$ and assuming \verb|Planck18| \citep{Planck_Cosmo_param_2018} as input cosmology. For simplicity, we assume direct knowledge of $\DM_{\rm diff}$, or in other words that one is capable of separating the other astrophysical components, e.g. the Milky Way or the host galaxy properties, affecting the FRBs' dispersion measure \citep[e.g.][]{FRB_DM_break_Thornton_et_al_2013, FRB_cosmo_DM_Deng_Zhang_2014, FRB_DM_break_Prochaska_Zheng_2019}. In reality, this is a stricter test for our approximation, since any errors in the inference of cosmological parameters can be directly traced to our approximation and not to other astrophysical effects that influence the $\DM$. 

We answer the following questions:

\begin{itemize}
    \item What happens when we use our approximation outside its region of best performance and applicability? (recall Section \ref{sec:accuracy_timing_performance})

    \item How does our approximation perform compared to the numerical solution with respect to timing in the cases of realistic cosmological analyses? (Section \ref{sec:cosmo_robust_time})

    \item What happens when we use our approximation with different PDFs for the diffuse dispersion measure of FRBs? (Section \ref{sec:cosmo_robust_time})

    \item As a concrete application, what happens when a different PDF for the Fast Radio Burst $\DM_{\rm diff}$ is used between data generation and cosmological inference? (Section \ref{sec:cosmo_pdf_modelling})
\end{itemize}

\begin{figure}[h]
    \centering
    \includegraphics[width=\linewidth]{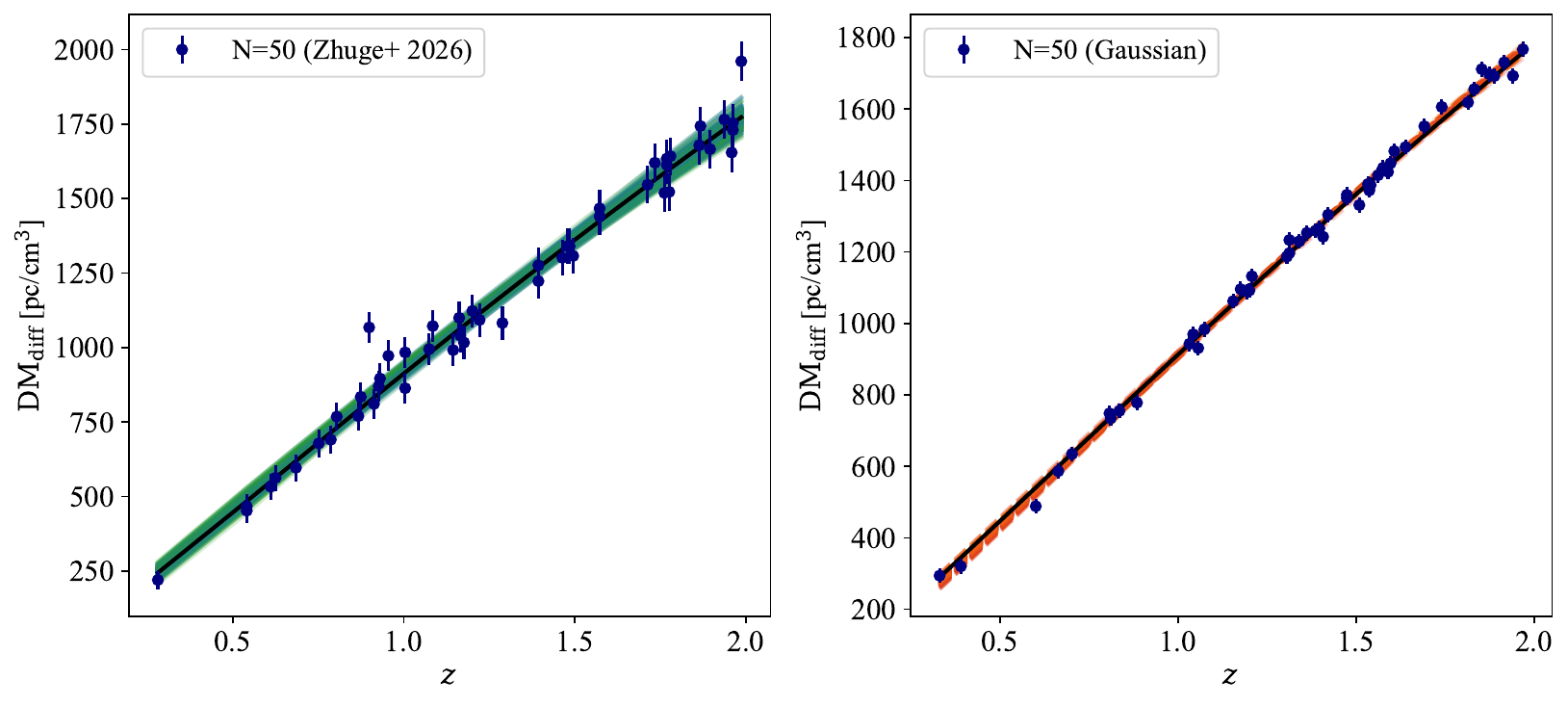}
    \caption{Simulated FRB signals following the two PDF choices (Left - \citep{FRBcosmo_localised_Zhuge_et_al_2026}, Right - Gaussian). In both plots, the blue dots show simulated events with their errors. The solid black line shows the fiducial cosmology. The coloured lines, follow the convention of Figure \ref{fig:fiducial_highz_joint_plot} (lower left panels), and represent a set of posterior predictive tests. The numerical integration and Pad\'e approximation lines are largely overlapping.}
    \label{fig:events_ppd_comparison_num_vs_pade}
\end{figure}

To generate a single FRB sample, we first draw a random redshift between $z=0.25$ and $z=2$ from a distribution\footnote{For our analysis, the choice of redshift distribution does not play a role. We just need a way to draw realistic redshift samples. Two example realisations are shown in Figure \ref{fig:events_ppd_comparison_num_vs_pade}.} that scales as the square of the comoving distance $D_c$ to the source $P(z) \sim D_c(z)^2$ \citep{Rate_distribution_Zhao_et_al_2011, Rate_distribution_Cai_Yang_2017}. For this redshift, we calculate the theoretical $\DM_{\rm diff}^{\rm th}$ from Eq. (\ref{eq:DM_diff}) and \verb|Planck18|. Then, we generate an observed $\DM_{\rm diff}^{\rm obs}$ by perturbing the theoretical value based on a PDF of choice (see below). We repeat $N=50$ times to obtain a series of ($z^{\rm obs}, \DM_{\rm diff}^{\rm obs}$) pairs.

For the PDF of $\DM_{\rm diff}$ we consider two cases: 

(1) A more realistic scenario that corresponds to the standard Macquart PDF \citep{McQuinn_2014, 2019_Prochaska_Zheng, Macquart+_2020} given by
\begin{equation}
    p_{\Delta, \rm Mac}(\Delta)={\cal A} \Delta^{-\beta}\exp\left[-\frac{(\Delta^{-\alpha}-C_0)^2}{2\alpha^2\sigma_{\rm diff, \Delta}^2}\right],
    \label{eq:p_Delta}
\end{equation}
where $\Delta={\DM_{\rm diff}}/\langle{\DM_{\rm diff}}\rangle$ is the dimensionless fraction between the measured and theoretical diffuse DM terms; ${\cal A}$ is the normalization factor for the PDF; $C_0$ is calculated by the condition that the mean value of $\Delta$ equals to $1$, which is true by definition; $\alpha$ and $\beta$ are parameters that describe the gas profile in cosmic halos and we adopt $\alpha=\beta=3$ \citep{Macquart+_2020}. Finally $\sigma_{\rm diff, \Delta}(S, z)$ is related to the scatter of $\Delta$ and we take the corrected form based on our previous paper \citep{FRBcosmo_localised_Zhuge_et_al_2026}.

(2) A simplified scenario, where the PDF is modelled by a Gaussian:

\begin{equation}
    p_{\rm Normal}(\DM_{\rm diff})=\frac{1}{\sqrt{2 \pi \sigma_{\rm diff}^2}}\exp\left[-\frac{(\DM_{\rm diff}-\DM_{\rm obs})^2}{2\sigma_{\rm diff, G}^2}\right],
    \label{eq:p_normal}
\end{equation}

with fiducial error $\sigma_{\rm diff, G} \sim 105$ pc/cm$^3$ \citep{Wei_et_al_FRB_GWs_Ass_2018}.

In both cases, we artificially reduce the errors to a value $\sigma_{\rm red} \sim (0.1-0.2) \times \sigma_{\rm diff, i}$, which results in tighter posteriors and where $i=\Delta$ or G depending on the PDF. This physically corresponds to detections with improved accuracy. This choice is motivated by our main focus on investigating the Pad\'e approximant in a relatively optimistic case, so that its potential limitations will not be masked by observational uncertainties.

\begin{table}[h]
    \centering
    \begin{tabular}{|c|c|c|}
    \hline
        Parameters & Range & Units\\
        \hline
        \hline
        $H_0$ & $\mathcal{U}(40, 100)$ & km/s/Mpc\\
        $\Omega_m$ & $\mathcal{U}(0, 1)$ & - \\
        $w$ & $\mathcal{U}(-2, 0)$ & - \\
    \hline
    \end{tabular}
    \caption{Cosmological parameters priors' distributions and range. $\mathcal{U}$ denotes a uniform distribution. We use the same priors both for the numerical integration case and for the Pad\'e approximant one. Note that for the latter, as shown in section \ref{sec:accuracy_timing_performance}, these correspond to regions of parameter space where our approximation underperforms, i.e. has fractional errors above $\Delta E \sim 1 \%$. We discuss the implications of this in the main text.}
    \label{tab:param_priors}
\end{table}

Finally, to obtain cosmological constraints we use a standard Bayesian framework. Starting from Bayes theorem, and ignoring the normalisation factor, we have

\begin{align}
    p( \mathbf{H} | \mathbf{D}, M) &\propto \mathcal{L}(\mathbf{D} | \mathbf{H}, M) \pi(\mathbf{H}) \nonumber \\
    & =  p_{\DM}(\DM_{\rm diff}^{\rm obs}, z^{\rm obs} | \DM_{\rm diff}^{\rm Pade}(\mathbf{H}, z), M) \pi(H_0) \pi(\Omega_m) \pi(w),
\end{align}

where for all the priors we assume uniform distributions (Table \ref{tab:param_priors}), $M$ denotes the model choice for $p_{\DM}$ (eqs. (\ref{eq:p_Delta}) and (\ref{eq:p_normal}), with the other FRB parameters fixed as described in the text) and $\mathbf{H}= \{H_0, \Omega_m, w\}$ and $\mathbf{D}= \{ z, \DM_{\rm diff}^{\rm obs} \}$ the cosmological and data vectors. We normalise the posterior at the end.

\subsection{Investigating robustness and timing}\label{sec:cosmo_robust_time}

\begin{figure}[h]
    \centering
        \includegraphics[width=\linewidth]{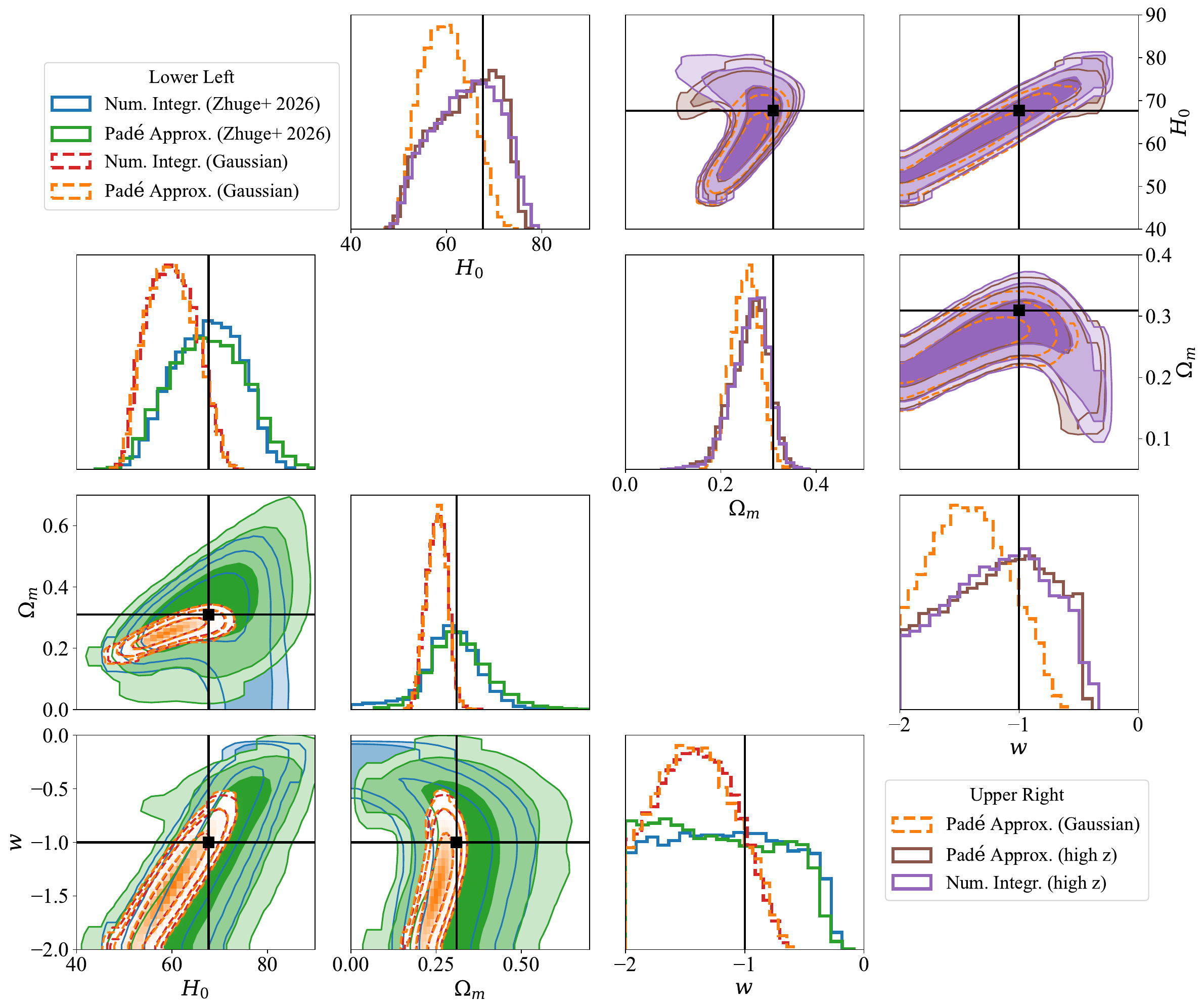}
        \caption{Cosmological inference of ($H_0, \Omega_m, w$) using numerical integration and our Pad\'e approximant for different modelling choices of the FRBs' $\DM_{\rm diff}$ PDF. (\emph{Lower Left}) We generate $N=50$ mock FRB $\DM_{\rm diff}$ using either the PDF developed in \citep{FRBcosmo_localised_Zhuge_et_al_2026} or a Gaussian, and we infer the cosmological parameters using either a numerical or an analytical approach. In all cases, the Pad\'e approximant gives identical results with the numerical ones, even in regions of parameter space where the approximation underperforms. (\emph{Upper Right}) Similar investigation for higher redshift events ($z>0.85$) - the two solid lines (numerical vs approximant) result to very similar constraints. The dashed line corresponds to a model misspecification investigation that is described in more detail in Section \ref{sec:cosmo_pdf_modelling}.}
        \label{fig:fiducial_highz_joint_plot}
\end{figure}

The results of the cosmological inference are collected in Figure \ref{fig:fiducial_highz_joint_plot}. In the lower left panels we are comparing the inference of cosmological parameters when using numerical integration vs our Pad\'e approximation, using different PDFs for the $\DM_{\rm diff}$ data generation and inference. Solid lines correspond to the model developed in \cite{FRBcosmo_localised_Zhuge_et_al_2026}, eq. (\ref{eq:p_Delta}), and dashed lines correspond to a Gaussian PDF model, eq. (\ref{eq:p_normal}). We observe that: 1) in all cases the input cosmological parameters are recovered. 2) Choice of PDF is not influencing the power of the approximation in any way, but note that as expected the more complex PDF yields wider constraints. 3) The numerical integration and Pad\'e approximants results are indistinguishable. This is true, even when including in our priors and inference regions of the parameter space where the fractional errors of the approximant become larger, i.e. above $\Delta E \sim 1 \%$. To quantify the comparison we report the posteriors median and $68 \%$ credible intervals (CI) for both the \cite{FRBcosmo_localised_Zhuge_et_al_2026} PDF and the Gaussian case in Table \ref{tab:posterior_ranges}.

In the upper right panels, we are repeating the analysis with $N=50$ events, but generating them only at higher redshifts ($0.85<z<2$). Distant sources have $\DM_{\rm diff}$ more susceptible to cosmological parameters, and as a consequence possible issues of the approximation could be revealed. In contrast, we find that our inference is equivalent independent of method used, i.e. numerical integration or Pad\'e approximation (solid lines). 

% Additionally, we investigated, anticipating the analysis of Section \ref{sec:cosmo_pdf_modelling}, the issue of PDF misspecification. The data were generated using the Gaussian PDF, and then inferred both using eqs. (\ref{eq:p_Delta}) - solid lines, and ((\ref{eq:p_normal}) - dashed lines. Although there exists a slight discrepancy in the 1D marginalised histograms, all models are consistent with each other and recovering the input values without any show of bias.

Finally, in Table \ref{tab:Timing_table_MCMC_Num_Pade}, we report the timing performance of our approximation in a full MCMC analysis. We note that in the case of the Gaussian PDF the timing improvement can be significant, reaching $\sim 27$ times faster inference. The improvement is more modest, about $3$ times faster computational times, when using a more complicated PDF, making the complexities of PDF modelling the more essential barrier.

\begin{table}[h]
    \centering
    \begin{tabular}{|c|c|c|c|c|}
    \hline
        Parameters$\backslash$PDF & Median (Num) & Median (Pad\'e) & $68 \%$ CI (Num) & $68 \%$ CI (Pad\'e)\\
        \hline
        \hline
        $H_0$ & $67.64$ & $67.49$ & $7.24$ & $8.24$ \\
        $\Omega_m$ & $0.3$ & $0.31$ & $0.07$ & $0.08$ \\
        $w$ & $-1.16$ & $-1.24$ & $0.57$ & $0.57$ \\
        \hline
        \hline
        Parameters$\backslash$Gauss & Median (Num) & Median (Pad\'e) & $68 \%$ CI (Num) & $68 \%$ CI (Pad\'e)\\
        \hline
        \hline
        $H_0$ & $59.91$ & $59.8$ & $5.21$ & $5.27$ \\
        $\Omega_m$ & $0.25$ & $0.25$ & $0.03$ & $0.03$ \\
        $w$ & $-1.41$ & $-1.42$ & $0.35$ & $0.35$ \\
    \hline
    \end{tabular}
    \caption{Posteriors medians and $68\%$ credible intervals in runs assuming two different PDF models (the more complex PDF from \cite{FRBcosmo_localised_Zhuge_et_al_2026} and a Gaussian). The numerical and Pad\'e results are practically equivalent.}
    \label{tab:posterior_ranges}
\end{table}

\begin{table}[h]
    \centering
    \begin{tabular}{|c|c|c|}
    \hline
        Model &  (PDF) (Num) & Gaussian (Num) \\
        \hline
        \hline
        (PDF) (Pad\'e) & $\sim 3$ & $-$ \\
        Gaussian (Pad\'e) & $-$ &  $\sim 27$ \\
    \hline
    \end{tabular}
    \caption{Timing improvement $\Delta t = t_{\rm Num}/t_{\rm App}$ values between the numerical integration and the Pad\'e approximants, within the $w$CDM model, in a full MCMC cosmological inference. We observe that the main bottleneck here is the modelling of the FRBs' $\DM$ PDF. For a simpler, Gaussian PDF the timing improvement is profound, while for a more precise, but computationally complex PDF the timing advantage is more modest.}
    \label{tab:Timing_table_MCMC_Num_Pade}
\end{table}

\begin{figure}[h]
    \centering
    \includegraphics[width=0.81\linewidth]{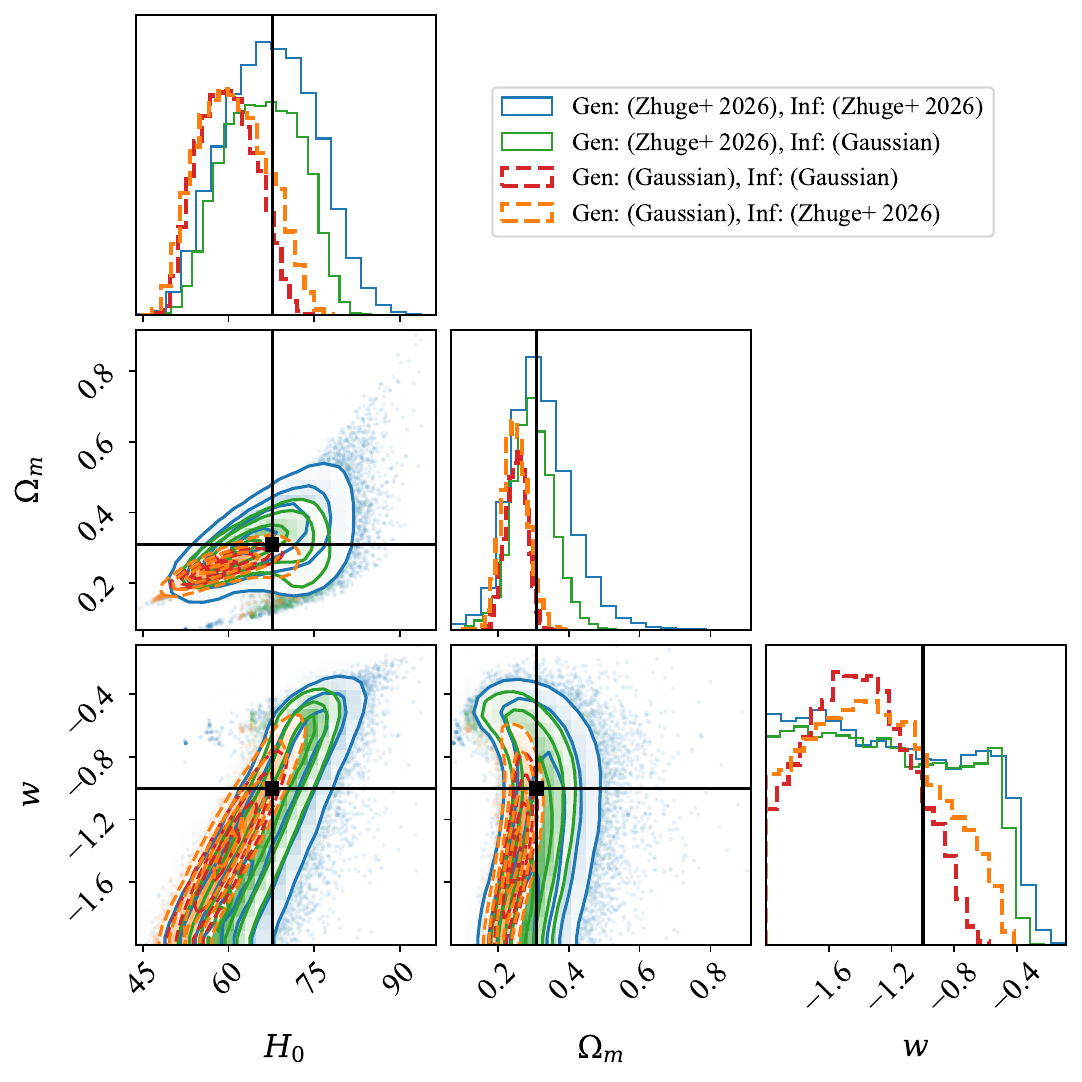}
    \caption{Cosmological inference of ($H_0, \Omega_m, w$) with our Pad\'e approximant for different modelling choices of the FRBs' $\DM_{\rm diff}$ PDF. We generate $N=50$ mock FRB $\DM_{\rm diff}$ using either the PDF developed in \citet{FRBcosmo_localised_Zhuge_et_al_2026} or a Gaussian, and we infer the cosmological parameters using the ``true'' and ``wrong'' PDFs, for four total configurations. We observe that in the case of a simple PDF (Gaussian - dashed lines) used in the data generation procedure, the PDF modelling employed in the cosmological inference is not affecting the result. In contrast, data generation through a more complex PDF (Eq. (\ref{eq:p_Delta}) - solid lines) requires similar modelling when inferring cosmological parameters to avoid biases.}
    \label{fig:MCMC_pdfs_comparison_gen_vs_inf}
\end{figure}

\subsection{Investigating PDF modelling}\label{sec:cosmo_pdf_modelling}

In this section we return in more detail to the final question: \emph{what happens when different PDF choices are used in data generation vs cosmological inference?}, as a concrete application of the Pad\'e approximants. In more astrophysical terms, what would be the effect of using a specific PDF modelling choice for FRBs $\DM_{\rm diff}$ which is not consistent with the actual distribution of $\DM_{\rm diff}$ of the Universe.

The results are summarised in Figure\footnote{In the caption, ``true'' refers to using for the cosmological inference the PDF used in the data generation, while ``wrong'' means that a different PDF is used in data generation vs cosmological inference.} \ref{fig:MCMC_pdfs_comparison_gen_vs_inf}. The analysis is using only the Pad\'e approximant for the inference, and $N=50$ events between $0.25<z<2$. The solid lines correspond to data generated through eq. (\ref{eq:p_Delta}), while the dashed lines through eq. (\ref{eq:p_normal}). We observe that in both cases, employing the same or different PDF in the inference of the cosmological parameters yields equivalent constraints. Additionally, we repeated the exercise for larger redshifts in the upper right panels of Figure \ref{fig:fiducial_highz_joint_plot}. The data were generated using the Gaussian PDF, and then inferred both using eqs. (\ref{eq:p_Delta}) - solid lines, and ((\ref{eq:p_normal}) - dashed lines. Although there exists a slight discrepancy in the 1D marginalised histograms, all models are consistent with each other and recovering the input values without any show of bias.

This is consistent with other findings in the literature \citep[e.g.][]{Konar_et_al_2025_Inference, Lemos_et_al_2026_FRB_DM_PDFs}, and points to the fact that in order for the details of the PDF modelling to become vital, more observational discriminative power will be needed. The only difference in the constraints lies in the data generation procedure, since the Gaussian PDF produce sources with smaller uncertainties. Consequently they lead to tighter posteriors.

\section{Conclusions}\label{sec:conclusion}

In this short paper, we have derived a Pad\'e approximant for the cosmic $\DM$ of FRBs, assuming a flat, $\Lambda$CDM or $w$CDM universe. We show that our approximation works well for a range of cosmological density parameters and redshifts. More specifically, for $0.01 \leq z \leq 2$, $0.2 \leq \Omega_m \leq 1.0$ and $-3.0 \leq w \leq -0.5$ the relative error in $\DM_{\rm diff}$ is always below $3.5 \%$. The relative error decreases as one probes larger redshifts $z > 0.01$, larger matter density $\Omega_m > 0.2$ and more negative DE EoS $w < -0.5$. 

Moreover, if one assumes the fiducial $\Lambda$CDM values for the pair $(\Omega_m, w) \simeq (0.31, -1)$, in order to constrain astrophysical and cosmological parameters in the factor outside the redshift integral $\DM_{\rm diff}^c$, the relative error is below $0.5 \%$ for the whole redshift range we considered. Compared to direct numerical evaluation, we find that our solution is more than $15$ times faster for $\Lambda$CDM and more than $2$ times faster for $w$CDM. 

In addition, we tested the approximation in two cosmologically relevant cases: First, a realistic scenario of cosmological inference using simulated FRB data in a $w$CDM universe. We showed that the numerical integration and the Pad\'e approximation are equivalent in all the cases we considered. Notably, we demonstrated that 1) the approximant provides similar cosmological constraints even when applied to regions of the parameters space where the relative error becomes more important, and 2) the performance is from $3$ to $27$ times faster, with the bigger bottleneck the specific modelling of the $DM_{\rm diff}$ PDF. Second, in a scenario where wrong modelling was used in cosmological inference vs data generation. We demonstrated that for current levels of observational accuracy the choice of modelling the PDF of the cosmic $\DM$ does not impact the cosmological constraints in a substantial way.

As a result, we consider our approximation a useful tool for cosmological and astrophysical studies involving FRBs.

\newpage

\paragraph{Software:} {\verb|Astropy| \citep{Astropy_2013, Astropy_2018}, \verb|Numpy| \citep{2020_Numpy}, \verb|Scipy| \citep{2020_SciPy}, \verb|Matplotlib| \citep{2007_Matplotlib}, \verb|Sympy| \citep{SymPy}. The code that reproduces our analysis can be found in \url{https://github.com/MariosNT/DM_cosmic_Pade}.}

\appendix

\vspace{-3mm}
\section{Comparison to the hypergeometric function}\label{App:Hypergeometric_function}

\cite{Jahns-Schindler_Spitler_2025} cite an analytical solution of the $\DM_{\rm diff}$ integral, for flat $\Lambda$CDM cosmologies with $\Omega_m < 1$ given by
\begin{equation}\label{eq:ap_hypergeometric_func}
    \int_0^z \frac{ (1+z) dz}{\sqrt{\Omega_m(1+z)^3+(1-\Omega_m)}} = \Bigg[ \frac{(z+1)^2}{2 \sqrt{1-\Omega_m}} \cdot \ _2F_1 \left(\frac{1}{2}, \frac{2}{3};\frac{5}{3};-\frac{\Omega_m (1+z)^3}{\Omega_\Lambda} \right) \Bigg] \Bigg|^z_0,
\end{equation}
with $_2F_1$ the hypergeometric function \citep{Abramowitz_Stegun_Math_functions}. This function is available in modern \verb|Python| packages, like \verb|Scipy| \citep{2020_SciPy}. Of course, our approximant is less precise in its range of validity compared to the analytic solution, however it is about $3$ times faster. This will give it an efficiency advantage when many $\DM$ calculations are needed, e.g. in the case of FRB cosmological inferences through MCMC analyses.

\vspace{-3mm}
\section{Coefficients of Pad\'e Approximant in $w$CDM}\label{App:Pade_wCDM_coeff}

As expected the coefficients are more complex this time, and to allow for better visibility they have been broken, where needed, in their numerator and denominator. In those cases, in eq. (\ref{eq:Pade_expansion_wCDM}), the coefficients are constructed as $b_i = b_i^{\rm num}/b_i^{\rm den}$ for example.

\begin{align}   
    c_1^{\rm num} = &7 \left(13554501120 w^{7} + 2144869632 w^{6} + 186662880 w^{5} + 11424240 w^{4} +\right. \nonumber \\ 
    &\left.  447552 w^{3} + 10512 w^{2} + 138 w + 1\right) \\ \nonumber \\
    c_1^{\rm den} = &4 \left(17420977152 w^{7} + 2348289792 w^{6} + 206452800 w^{5} + 12398832 w^{4} +\right. \nonumber \\
    &\left. 473904 w^{3} + 10800 w^{2} + 144 w + 1\right) \\ \nonumber \\
    c_2^{\rm num} = &7 \left(287698065408 w^{8} + 67349242368 w^{7} + 7092738432 w^{6} + 484856928 w^{5} + \right. \nonumber \\ 
    &\left.  23238576 w^{4} + 732672 w^{3} + 14328 w^{2} + 162 w + 1 \right) \\ \nonumber \\
    c_2^{\rm den} = &8 \left(522629314560 w^{8} + 87869670912 w^{7} + 8541873792 w^{6} + 578417760 w^{5} + \right. \nonumber \\
    &\left. 26615952 w^{4} + 797904 w^{3} + 15120 w^{2} + 174 w + 1\right) \\ \nonumber \\
    c_3^{\rm num} = &7 \left(3862337458176 w^{9} + 1485633788928 w^{8} + 219003450624 w^{7} + 17781674688 w^{6} + \right. \nonumber \\ 
    &\left. 976307904 w^{5} + 38712816 w^{4} + 1042416 w^{3} + 17892 w^{2} + 180 w + 1 \right) \\ \nonumber \\
    c_3^{\rm den} = &64 \left(12543103549440 w^{9} + 2631501416448 w^{8} + 292874641920 w^{7} + 22423900032 w^{6} + \right. \nonumber \\
    &\left. 1217200608 w^{5} + 45765648 w^{4} + 1160784 w^{3} + 19296 w^{2} + 198 w + 1\right) \\ \nonumber \\
    b_0 = &6 w \\ \nonumber \\
    b_1^{\rm num} = &15 w \left(113857809408 w^{8} + 30024815616 w^{7} + 3631469760 w^{6} + 274710528 w^{5} + \right. \nonumber \\ 
    &\left.  14793840 w^{4} + 525312 w^{3} + 11556 w^{2} + 144 w + 1 \right) \\ \nonumber \\
    b_1^{\rm den} = &2 \left(104525862912 w^{8} + 31510715904 w^{7} + 3587006592 w^{6} + 280845792 w^{5} + \right. \nonumber \\
    &\left. 15242256 w^{4} + 538704 w^{3} + 11664 w^{2} + 150 w + 1\right) \\ \nonumber \\
    b_2^{\rm num} = &9 w \left(48333274988544 w^{10} + 15147986411520 w^{9} + 2611359461376 w^{8} + \right. \nonumber \\ 
    &\left. 276998911488 w^{7} + 19936155456 w^{6} + 1037489472 w^{5} + 39459312 w^{4} + \right. \nonumber \\
    &\left. 1036800 w^{3} + 17676 w^{2} + 180 w + 1 \right) \\ \nonumber \\
    b_2^{\rm den} = &4 \left(37629310648320 w^{10} + 15733943967744 w^{9} + 2719298304000 w^{8} + \right. \nonumber \\
    &\left. 283269477888 w^{7} + 20869742016 w^{6} + 1114953984 w^{5} + 42066864 w^{4} + \right. \nonumber \\
    &\left. 1082592 w^{3} + 18324 w^{2} + 192 w + 1\right) \\ \nonumber \\
    b_3^{\rm num} = &3 w \left(35039125420572672 w^{12} + 8629966763753472 w^{11} + 1619918846263296 w^{10} +  \right. \nonumber \\ 
    &\left. 221082842105856 w^{9} + 21477950191872 w^{8} + 1497341832192 w^{7} + 77109117696 w^{6} + \right. \nonumber \\
    &\left. 3001784832 w^{5} + 87399648 w^{4} + 1812672 w^{3} + 25200 w^{2} + 216 w + 1 \right) \\ \nonumber \\
    b_3^{\rm den} = &32 \left(16255862200074240 w^{12} + 8377494841294848 w^{11} + 1873191824621568 w^{10} +  \right. \nonumber \\
    &\left. 252316887183360 w^{9} + 23632344926208 w^{8} + 1641458763648 w^{7} + 85870694592 w^{6} + \right. \nonumber \\
    &\left. 3349442016 w^{5} + 95451696 w^{4} + 1935144 w^{3} + 26820 w^{2} + 234 w + 1\right)     
\end{align}

\bibliographystyle{mnras}
\bibliography{biblio.bib} 

\end{document}